\documentclass[conference,10pt]{IEEEtran}
\usepackage{epsfig,rotating,setspace,latexsym,amsmath,epsf,amssymb,amsfonts,bm,theorem,subfigure,epstopdf}
\usepackage{cite,authblk}
\usepackage{bbm}
\usepackage[ruled,vlined]{algorithm2e}

\usepackage{color}

\DeclareMathOperator*{\argminA}{arg\,min}

\SetArgSty{textnormal}

\IEEEoverridecommandlockouts
\allowdisplaybreaks

\begin{document}
\bstctlcite{IEEEexample:BSTcontrol} 

\title{Adversarial Attacks on Deep Learning Based mmWave Beam Prediction in 5G and Beyond}
	
\author[1]{Brian Kim}
\author[2]{Yalin E. Sagduyu}
\author[2]{Tugba Erpek}
\author[1]{Sennur Ulukus}

\affil[1]{\normalsize Department of Electrical and Computer Engineering, University of Maryland, College Park, MD 20742, USA}
\affil[2]{\normalsize Intelligent Automation, Inc., Rockville, MD 20855, USA  \thanks{This effort is supported by the U.S. Army Research Office under contract W911NF-20-C-0055. The content of the information does not necessarily reflect the position or the policy of the U.S. Government, and no official endorsement should be inferred.}}
	
\maketitle

\begin{abstract}
Deep learning provides powerful means to learn from spectrum data and solve complex tasks in 5G and beyond such as beam selection for initial access (IA) in mmWave communications. To establish the IA between the base station (e.g., gNodeB) and user equipment (UE) for directional transmissions, a deep neural network (DNN) can predict the beam that is best slanted to each UE by using the received signal strengths (RSSs) from a subset of possible narrow beams. While improving the latency and reliability of beam selection compared to the conventional IA that sweeps all beams, the DNN itself is susceptible to adversarial attacks. We present an adversarial attack by generating adversarial perturbations to manipulate the over-the-air captured RSSs as the input to the DNN. This attack reduces the IA performance significantly and fools the DNN into choosing the beams with small RSSs compared to jamming attacks with Gaussian or uniform noise.
\end{abstract}

\section{Introduction}\label{sec:Introduction}

Due to the algorithmic and computational advances in deep learning (DL), various applications in different domains, such as computer vision \cite{vision1} and speech recognition \cite{speech1}, have been empowered by deep neural networks (DNNs) in solving complex problems by effectively learning from rich data representations. Similarly, DL has been applied for various wireless communication tasks, such as waveform design, signal classification, spectrum sensing, and interference management \cite{erpek1} by capturing the intrinsic characteristics of the spectrum data.

One particular application of DL is in the domain of initial access (IA), where user equipments (UEs) need to establish their initial connection to an access point or base station when they attempt to join the communication network for the first time \cite{giordani2016initial}. As 5G and beyond communication systems rely on millimeter wave (mmWave) and higher frequency bands to sustain high data rates over large available bandwidths, transmissions become more directional using narrow beams. This paradigm makes the beam alignment in the IA process more difficult as many narrow beams need to be swept to find the most suitable beam for each UE \cite{Li,Alkhateeb}. In the IA, the transmitter such as the 5G base station (gNodeB) sequentially transmits pilot signals over different narrow beams. The UE calculates the received signal strength (RSS) for each beam, determines the beam that provides the highest RSS, and informs the gNodeB of this beam selection. Since the time for the IA is limited, it is essential to reduce the number of beams swept as checking each beam consumes time and delays the UE to gain access to time-sensitive services such as ultra-reliable low-latency communications (URLLC) in 5G. To overcome this issue, a DL-based approach has been proposed in \cite{Cousik, Cousik2} to reduce the number of beams that need to be swept before making a decision, compared to the conventional beam sweeping approach that needs to sweep all possible beams. In this DL-based approach, the UE predicts the best beam from a large set of narrow beams by using only the RSSs for a subset of these possible beams. 

It is well known that DNNs are highly susceptible to carefully crafted adversarial perturbations that induce incorrect output or misclassification, as first shown in computer vision applications \cite{Szegedy1}. Wireless medium is shared and open to jamming attacks that can also be launched via adversarial machine learning to target the underlying DNNs. Therefore, adversarial machine learning has recently gained attention as the emerging attack surface for wireless security \cite{Sagduyu2020}. The attacks built upon adversarial machine learning include exploratory (inference) attacks \cite{Shi2018AdDL4CogRaSec, Terpek}, adversarial (evasion) attacks \cite{Larsson2, Kokalj2, Kokalj3, Flowers1, Bair1, Lin, Kim1, Kim2, Gunduz2, Gunduz1, Kim5G, KimMultiple, KimICC, IoT}, poisoning (causative) attacks \cite{YiMilcom2018,Sagduyu1, Luo2019, ZluoPartialAttack}, membership inference attacks \cite{MIA}, Trojan attacks \cite{Davaslioglu1}, and spoofing attacks \cite{Shi2019generative, ShiGANSpoofing}. These attacks are stealthier (harder to detect) than conventional jamming schemes \cite{Sagduyu2008, Sagduyuuncertainty}.  

In this paper, we consider an adversarial attack that aims to manipulate the input to a DNN in test time. Most of the applications of adversarial attacks to the wireless domain have focused on wireless signal classifiers such as modulation classifiers \cite{Larsson2, Kokalj2, Kokalj3, Flowers1, Bair1, Lin, Kim1, Kim2, Gunduz2, Gunduz1, Kim5G, KimMultiple, KimICC} and spectrum sensing classifiers \cite{IoT,YiMilcom2018,Sagduyu1}. There have been efforts to extend adversarial attacks to communication problems such as autoencoder-based end-to-end communications \cite{Larsson1}, power control for MIMO communications \cite{Larsson3}, and dynamic channel access \cite{zhong2020adversarial, wang2020defense}. In addition, adversarial machine learning has been used to develop attacks on 5G-specific tasks such as 5G spectrum sharing with incumbent users \cite{sagduyubc}, 5G signal authentication \cite{sagduyubc}, and 5G radio access network (RAN) slicing \cite{shiRL, shiflooding}.

Our goal in this paper is to investigate the vulnerability of a DNN that is used for mmWave beam prediction as part of the IA process in 5G and beyond communications. We first generate a non-targeted attack using fast gradient method (FGM) \cite{Kurakin1} to cause any misclassification at the DNN classifier at the receiver (independent of the wrong labels for beams). Then, we introduce the $k$-worst beam attack that not only causes misclassification but also enforces the DNN classifier to select one of the $k$ worst beams to further reduce the IA performance. We compare the non-targeted FGM attack and $k$-worst beam attack with benchmark jamming attacks with Gaussian and uniform noise added across RSSs from input beams. Our results show that the beam prediction of the IA process is highly vulnerable to adversarial attacks that can significantly reduce the beam prediction accuracy. The effect of this attack translates to notable reduction in communication performance in terms of data rate since the UEs connect to the gNodeB using beams that are not well aligned and the corresponding RSSs drop significantly.   

The rest of the paper is organized as follows. Section~\ref{sec:SystemModel} provides the system model.  Section~\ref{sec:attacks} describes the adversarial attacks considered in this paper, namely, the non-targeted attack and the $k$-worst attack. Section~\ref{sec:simulation} presents the performance results. Section~\ref{sec:Conclusion} concludes the paper.

\section{System Model} \label{sec:SystemModel}

We consider a mmWave network that consists of a directional transmitter (e.g., the gNodeB in 5G), an omnidirectional receiver (e.g., a UE) during the IA, and an adversary. A pre-trained DL-based classifier is applied to the RSS values measured at the receiver to select the best beam without a need to sweep all beams, as proposed in \cite{Cousik, Cousik2}. Concurrently, the adversary attempts to cause misclassification at the receiver by launching an over-the-air adversarial attack.

\begin{figure}[t]
	\centerline{\includegraphics[width=0.925\linewidth]{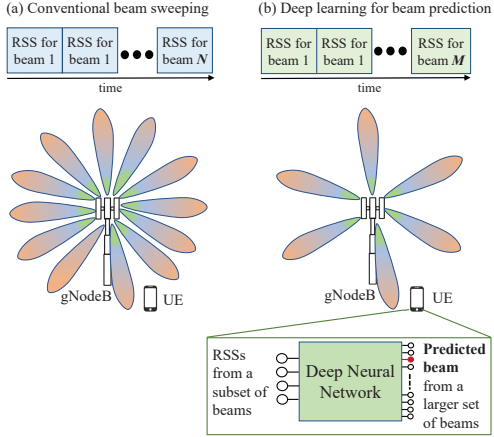}}
	\caption{Conventional beam sweeping vs. DL-based beam prediction.}
	\label{plot:IA}
\end{figure}

Fig.~\ref{plot:IA} illustrates two  approaches for the IA, namely conventional beam sweeping and DL for beam prediction. In the conventional IA process with exhaustive beam sweeping, the transmitter transmits pilot signals over all narrow beams and the receiver uses the RSS values from all $\mathcal{N}$ possible beams where $N = |\mathcal{N}|$ to select the best beam. The transmitter transmits the pilot signal over each beam at a separate time slot. DL reduces the number of beams swept such that the DNN classifier at the receiver only uses the RSS values from $\mathcal{M}$ subset of beams, where $\mathcal{M} \subseteq \mathcal{N}$ and $M = |\mathcal{M}|$, to select the best beam. For that purpose, the transmitter transmits the pilot signals over a smaller set of beams, each at a separate time slot. Overall, the total time needed for the IA is reduced. In the DL-based approach, the DNN classifier at the receiver is denoted by $f(.;\boldsymbol{\theta}): \mathcal{X}_{M} \rightarrow \mathbb{R}^{N}$ where $\mathcal{X}_{M} \subset \mathbb{R}^{M}$ is the input to the DNN which corresponds to the RSS values of $M$ beams and $\boldsymbol{\theta}$ is the set of the receiver's DNN parameters. The classifier $f$ assigns the best beam $\hat{l}(\boldsymbol{x}_M,\boldsymbol{\theta}) = \arg \max_{j} f_{j}(\boldsymbol{x}_M,\boldsymbol{\theta})$ to every input $\boldsymbol{x}_M\in \mathcal{X}_{M}$. In this formulation, $f_{j}(\boldsymbol{x}_M,\boldsymbol{\theta})$ is the output of classifier $f$ corresponding to the $j$th beam.

\begin{figure}[t]
 	\centerline{\includegraphics[width=0.925\linewidth]{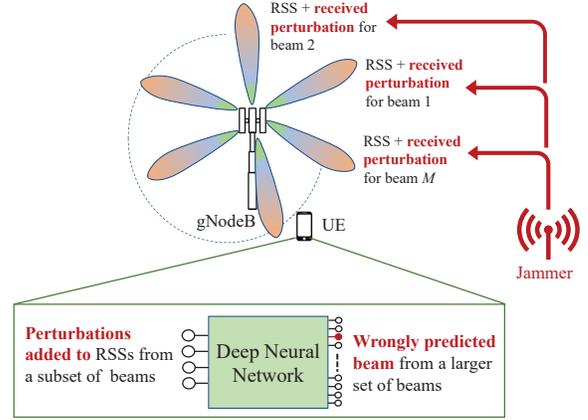}}
 	\caption{Adversarial attack on beam prediction with deep learning.}
 	\label{plot:IA2}
\end{figure}

In the meantime, as shown in Fig.~\ref{plot:IA2}, the adversary generates an adversarial attack by jamming the spectrum with adversarial perturbation, $\boldsymbol{\delta} = [\delta_{1}, \delta_{2},\cdots,\delta_{M}]$, under some suitable power constraint $P_{\textit{max}}$ with respect to the original RSS values, $\boldsymbol{x}\in \mathbb{R}^{N}$, by solving the following optimization problem:
\begin{align} \label{eq:perturbation}
	\argminA_{\boldsymbol{\delta}}& \quad \sum_{i=1}^{M}\delta_{i}\nonumber\\
	s.t. &\quad\hat{l}(\boldsymbol{x}_M,\boldsymbol{\theta}) \ne \hat{l}(\boldsymbol{x}_M+\boldsymbol{\delta},\boldsymbol{\theta}) \nonumber \\ &\quad  \sum_{i=1}^{M}{\delta_{i}} \le P_{\textit{max}}. 
\end{align}
Note that $\boldsymbol{x}_M\in \mathbb{R}^{M}$ is the $M$ elements from $\boldsymbol{x}$ which correspond to the RSS values at the receiver for all beams. 
 
However, solving (\ref{eq:perturbation}) is hard due to the inherent structure of the DNN. Thus, different methods have been proposed to approximate the adversarial attack such as FGM. FGM is a computationally efficient way of generating adversarial attacks by linearizing the loss function of the DNN classifier. We denote the loss function of the DNN classifier by $L(\boldsymbol{\delta},\boldsymbol{x},\boldsymbol{y})$, where $\boldsymbol{y}\in \{0,1\}^{N}$ is the one-hot encoded class vector. Then, FGM linearizes this loss function in a neighborhood of $\boldsymbol{x}$ and uses this linearized function for optimization. 
 
Throughout the paper, we assume that the adversary knows the architecture ($\boldsymbol{\theta}$ and $L(\cdot)$) of the DNN classifier at the receiver, and the exact RSS values for all $N$ beams that are received at the receiver. Later in the paper, we will relax the assumption about knowing all $N$ RSS values to knowing only $M$ RSS values, and compare the results under these two assumptions. In the next section, we describe the two adversarial attacks considered in this paper, namely, the non-targeted attack and the $k$-worst attack.

\section{Adversarial Attacks on Beam Prediction} \label{sec:attacks}
\subsection{Non-targeted FGM Attack}
First, we consider a non-targeted FGM attack where the adversary searches for a perturbation that causes any misclassification at the receiver's DNN classifier. For that purpose, the adversary designs a perturbation that maximizes the loss function $L(\boldsymbol{\delta},\boldsymbol{x}_{M},\boldsymbol{y}^{\textit{true}})$, where $\boldsymbol{y}^{\textit{true}}$ is the true label of $\boldsymbol{x}_{M}$. FGM is used to linearize the loss function as $L(\boldsymbol{\theta},\boldsymbol{x}_{M}+\boldsymbol{\delta},\boldsymbol{y}^{\textit{true}}) \approx L(\boldsymbol{\theta},\boldsymbol{x}_{M},\boldsymbol{y}^{\textit{true}}) + \boldsymbol{\delta}^{T} \nabla_{\boldsymbol{x}_{M}}L(\boldsymbol{\theta},\boldsymbol{x}_{M},\boldsymbol{y}^{\textit{true}})$ that is maximized by setting $\boldsymbol{\delta} = \alpha \nabla_{\boldsymbol{x}_{M}}L(\boldsymbol{\theta},\boldsymbol{x}_{M},\boldsymbol{y}^{\textit{true}})$, where $\alpha$ is a scaling factor to constrain the adversarial perturbation power to $P_{\textit{max}}$. The details of selecting $\alpha$ and generating the non-targeted FGM attack are presented in Algorithm \ref{non-target}.

\begin{algorithm}[t]
	\DontPrintSemicolon
	\SetAlgoLined
	\label{non-target}
	Inputs: RSS values $\boldsymbol{x}_M$, true label $\boldsymbol{y}^{\textit{true}}$, desired accuracy $\varepsilon_{acc}$, power constraint $P_{\textit{max}}$, and model of the classifier \\
	Initialize: ${\varepsilon}\leftarrow {0}, \varepsilon_{\textit{max}} \leftarrow {P_{\textit{max}}}, \varepsilon_{min} \leftarrow 0$\\
	$\boldsymbol{\delta}_{norm} =\frac{\nabla_{\boldsymbol{x}_M}L(\boldsymbol{\theta},\boldsymbol{x}_{M},\boldsymbol{y}^{\textit{true}})}{(||\nabla_{\boldsymbol{x}_M}L(\boldsymbol{\theta},\boldsymbol{x}_{M},\boldsymbol{y}^{\textit{true}})||_{1})}$\\
	\If{$\hat{l}(\boldsymbol{x}_{M},\boldsymbol{\theta})== \boldsymbol{y}^{\textit{true}}$}{
	\While{$\varepsilon_{\textit{max}}-\varepsilon_{min} > \varepsilon_{acc}$}{
		$\varepsilon_{avg} \leftarrow (\varepsilon_{\textit{max}}+\varepsilon_{min})/2$\\
		$\boldsymbol{x}_{adv} \leftarrow \boldsymbol{x}_{M} + \varepsilon_{avg}\boldsymbol{\delta}_{\textit{norm}}$\\
		\lIf{$\hat{l}_{m}(\boldsymbol{x}_{adv})== \boldsymbol{y}^{\textit{true}}$}{
			$\varepsilon_{min}\leftarrow \varepsilon_{avg}$}
		\lElse{$\varepsilon_{\textit{max}}\leftarrow \varepsilon_{avg}$}
	}}{$\varepsilon = \varepsilon_{\textit{max}}$, $\boldsymbol{\delta}^{*} = \varepsilon\boldsymbol{\delta}_{norm} $}\\
	\caption{Non-targeted FGM attack}
\end{algorithm}

\subsection{$k$-worst Beam Attack}
Next, the adversary designs an adversarial attack such that it not only causes a misclassification at the receiver's DNN classifier but also tries to change the beam to one of the worst $k$ beams. Unlike attacks on some signal classification tasks (in the computer vision or wireless domains), where changing the label from `signal 1' to `signal 2' and from `signal 1' to `signal 3' may have the same effect on the classifier meaning that both correspond to an error in classification, changing the label from the best beam to second worst beam and from the best beam to the worst beam have totally different effects on the communication performance (as the signal quality on a beam strongly affects the achieved rate following the IA). Thus, the adversary first tries to change the label to the worst beam and if the adversary is not able to change to that label, then the adversary tries to change the label to the second worst beam. The adversary continues to do so until it tries for the $k$th worst beam. 

Here, we consider the targeted FGM attack where the adversary aims to fool the DNN to a target label. Therefore, the adversary tries to minimize the loss function $L(\boldsymbol{\delta},\boldsymbol{x}_M,\boldsymbol{y}^{\textit{target}})$, where $\boldsymbol{y}^{\textit{target}}$ is one of the $k$ worst beams. FGM is used again to linearize the loss function 
as $L(\boldsymbol{\theta},\boldsymbol{x}_{M}+\boldsymbol{\delta},\boldsymbol{y}^{\textit{target}}) \approx L(\boldsymbol{\theta},\boldsymbol{x}_{M},\boldsymbol{y}^{\textit{target}}) + \boldsymbol{\delta}^{T} \nabla_{\boldsymbol{x}_{M}}L(\boldsymbol{\theta},\boldsymbol{x}_{M},\boldsymbol{y}^{\textit{target}})$ that is minimized by setting $\boldsymbol{\delta} = -\alpha \nabla_{\boldsymbol{x}_{M}}L(\boldsymbol{\theta},\boldsymbol{x}_{M},\boldsymbol{y}^{\textit{target}})$, where $\alpha$ is a scaling factor to constrain the adversarial perturbation power to $P_{\textit{max}}$. First, we assume that the adversary knows the order of beam indices based on the real RSS values. Then, we relax this assumption by getting the order from the output of the DNN classifier. Since the DNN classifier is only trained to find the best beam, there exists a discrepancy between the real order of beams and the order of beams obtained from the DNN output. The details of the algorithm are presented in Algorithm \ref{q worst}.

\begin{algorithm}[t]
	\DontPrintSemicolon
	\SetAlgoLined
	\label{q worst}
	Inputs: RSS values $\boldsymbol{x}_M$, true label $\boldsymbol{y}^{\textit{true}}$, $k$ worst beam indices, desired accuracy $\varepsilon_{acc}$, power constraint $P_{\textit{max}}$, and model of the classifier \\
	\If{$\hat{l}(\boldsymbol{x}_{M},\boldsymbol{\theta})== \boldsymbol{y}^{\textit{true}}$}{
	\For{$i$ in $k$ worst beam indices}{
	Initialize: ${\varepsilon}\leftarrow {0}, \varepsilon_{\textit{max}} \leftarrow {P_{\textit{max}}}, \varepsilon_{min} \leftarrow 0, \boldsymbol{y}^{\textit{target}}\leftarrow i$\\
	$\boldsymbol{\delta}_{norm} =\frac{\nabla_{\boldsymbol{x}_M}L(\boldsymbol{\theta},\boldsymbol{x}_{M},\boldsymbol{y}^{\textit{target}})}{(||\nabla_{\boldsymbol{x}_M}L(\boldsymbol{\theta},\boldsymbol{x}_{M},\boldsymbol{y}^{\textit{target}})||_{1})}$\\
	
	\While{$\varepsilon_{\textit{max}}-\varepsilon_{min} > \varepsilon_{acc}$}{
		$\varepsilon_{avg} \leftarrow (\varepsilon_{\textit{max}}+\varepsilon_{min})/2$\\
		$\boldsymbol{x}_{adv} \leftarrow \boldsymbol{x}_{M} - \varepsilon_{avg}\boldsymbol{\delta}_{\textit{norm}}$\\
		\lIf{$\hat{l}_{m}(\boldsymbol{x}_{adv})== \boldsymbol{y}^{\textit{true}}$}{
			$\varepsilon_{min}\leftarrow \varepsilon_{avg}$}
		\lElse{$\varepsilon_{\textit{max}}\leftarrow \varepsilon_{avg}$}
	}
	$\boldsymbol{x}_{adv} \leftarrow \boldsymbol{x}_{M} - \varepsilon_{avg}\boldsymbol{\delta}_{\textit{norm}}$\\
	\lIf{$\hat{l}_{m}(\boldsymbol{x}_{adv})== i$}{break}
	}}
	{$\boldsymbol{\delta}^{*} = -\varepsilon_{avg}\boldsymbol{\delta}_{norm} $}\\
	\caption{$k$-worst beam attack}
\end{algorithm}

\section{Performance Evaluation}\label{sec:simulation}
This section describes the details of the DNN classifier used at the receiver and the dataset that is used to train it. Furthermore, the performances of the two attack schemes that are introduced in this paper are compared with two benchmark jamming attacks that inject Gaussian or uniform noise on different beams.

\subsection{Deep Learning Framework for IA}
We use the DNN structure that consists of seven layers including the input and the output layers as shown in Table~\ref{table1}. The input layer has $M =12$ neurons and the output layer has $N=24$ neurons where each neuron represents a likelihood score for each beam. Note that each dense layer's output is batch normalized before sending it to the next layer. We follow the simulation setup of \cite{Cousik} and apply the attacks in a two-dimensional mmWave network scenario where line-of-sight (LoS) mmWave channels with pathloss and shadowing effects are considered. A $10 \times 10$ antenna array is used at the transmitter to generate a beam width of about $15 ^{\circ}$ using the standard planar array formulation. The location of the transmitter is fixed at $(0,0)$ and the receiver's $x$ and $y$ positions are each uniformly randomly distributed between $-25$ m and $25$ m. This bounds the simulation cell to an area of $50 \times 50$ m. The transmit power is set to $20$ dBm. A total of $10^6$ receiver positions are generated, which act as data samples. The adversary is also uniformly distributed in this area and generates perturbations that are added to the data samples.

\begin{table}[t]
\caption{The DNN architecture for the IA.}
\begin{tabular}{l|l|l}
 Layers&Number of neurons  & Activation function  \\ \hline
Input & $M=12$ & - \\ \hline
 Dense 1& 32 & ReLu \\ \hline
Dense 2&  64& ReLu \\ \hline
 Dense 3& 126 & ReLu \\ \hline
 Dense 4& 64 & ReLu \\ \hline
 Dense 5& 32 &  ReLu\\ \hline
 Output&  $N=24$& Softmax
\end{tabular}\label{table1}
\end{table}

\subsection{Attack Performance Results}
We compare the two attack schemes that we have described earlier with the Gaussian attack and the uniform attack that generate perturbations with Gaussian and equal power distribution for $M$ beams, respectively. In the simulations, we use the perturbation-to-signal ratio (PSR) metric that shows the relative perturbation power with respect to the received signal power (namely, the RSS at the receiver). As the PSR increases, the attack becomes more likely to be detected.

\begin{figure}[t]
	\centerline{\includegraphics[width=0.905\linewidth]{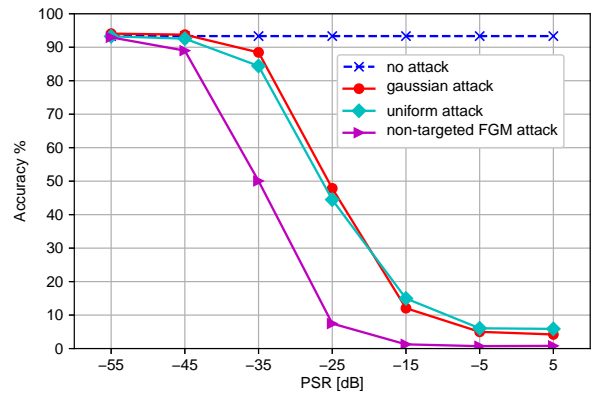}}
	\caption{Classifier accuracy under the non-targeted attack.}
	\label{plot:nontarget}
\end{figure}

Fig.~\ref{plot:nontarget} presents the accuracy of the classifier at the receiver under the non-targeted FGM attack and compares it with the Gaussian attack and the uniform attack. The non-targeted FGM attack significantly impacts the accuracy of the classifier for beam selection even for signal strength fluctuations that cannot be resolved with typical hardware. Also, the Gaussian attack and the uniform attack do not perform well compared to the non-targeted attack in [-40dB,-20dB] region, while they are comparable in the high PSR region. 

\begin{figure}[t]
	\centerline{\includegraphics[width=0.905\linewidth]{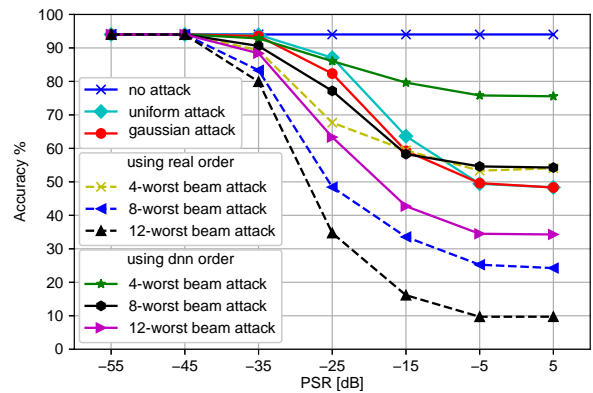}}
	\caption{Classifier accuracy under the $k$-worst beam attack.}
	\label{plot:qworst}
\end{figure}

In Fig.~\ref{plot:qworst}, we investigate the performance of the $k$-worst beam attack, where $k$ is set as 4, 8, 12, and compare it with the Gaussian attack and the uniform attack. Note that the accuracy definition in Fig.~\ref{plot:qworst} is different from Fig.~\ref{plot:nontarget}, where the accuracy is defined as the percentage that the label obtained from the DNN classifier is in the $N-k$ best beams since the attack is successful only if the $k$-worst beam attack fools the DNN classifier into choosing one of the worst $k$ beams. As $k$ increases for the $k$-worst beam attack, the DNN classifier accuracy decreases meaning that it is harder to enforce the beam selection to the worst ones. Also, the $k$-worst beam attack using the real order of RSSs outperforms the $k$-worst beam attack using the DNN order as there is a discrepancy between the real order and the DNN-predicted order of RSSs. Furthermore, the Gaussian attack and the uniform attack both saturate around 50\% meaning that under both attacks the beams are misclassified to the best or worst group 50\% of the time.

\section{Conclusion} \label{sec:Conclusion}

We presented an adversarial attack to fool the beam selection process of the IA using a DNN classifier that uses a subset of beams to predict the beam that is best oriented to the receiver. We investigated two different attack methods, namely, the non-targeted FGM attack that only aims to fool the DNN classifier with misclassification to any other beam label, and the $k$-worst beam attack that not only fools the DNN classifier but also enforces the label that is chosen at the DNN to be in one of the $k$-worst beams. We showed that the adversarial attack can significantly decrease the accuracy of the DNN and fool the DNN into selecting the worst beam. Results demonstrate that as DL finds applications for beam prediction in mmWave communication for 5G and beyond, the IA process becomes vulnerable to adversarial attacks that can significantly reduce the beam selection performance.     

\newpage
\bibliographystyle{IEEEtran}
\bibliography{lib}

\end{document}